# The European research elite: a cross-national study of highly productive academics in 11 countries

Marek Kwiek[1]

**Abstract** In this paper, we focus on a rare scholarly theme of highly productive academics, statistically confirming their pivotal role in knowledge production across 11 systems studied. The upper 10 % of highly productive academics in 11 European countries studied ($N = 17,211$) provide on average almost half of all academic knowledge production. In contrast to dominating bibliometric studies of research productivity, we focus on academic attitudes, behaviors, and perceptions as predictors of becoming research top performers across European systems. Our paper provides a (large-scale and cross-country) corroboration of the systematic inequality in knowledge production, for the first time argued for by Lotka (J Wash Acad Sci 16:317–323, 1929) and de Solla Price (Little science, big science. Columbia University Press, New York, 1963). We corroborate the deep academic inequality in science and explore this segment of the academic profession. The European research elite is a highly homogeneous group of academics whose high research performance is driven by structurally similar factors, mostly individual rather than institutional. Highly productive academics are similar from a cross-national perspective, and they substantially differ intra-nationally from their lower-performing colleagues.

✉ Marek Kwiek
kwiekm@amu.edu.pl

[1] Center for Public Policy Studies, UNESCO Chair in Institutional Research and Higher Education Policy, University of Poznan, Ul. Szamarzewskiego 89, 60-569 Poznan, Poland

# Introduction

This paper focuses on a unique class of highly productive academics in Europe, as well as on the predictors for becoming highly productive, from a European cross-national comparative perspective. The paper starts with an empirical finding that 10 % of academics provide on average almost half of all academic knowledge production in the 11 countries studied. Beginning with the remarkable similar productivity distribution patterns across European systems, we pose a general research question: Who are these highly productive academics and which institutional and/or individual factors increase the odds of entering this class?

Highly productive academics as a separate segment of the academic profession are a very rare scholarly theme. Following a handful of previous studies focusing on the theme to varying degrees (de Solla Price 1963; Crane 1965; Prpić 1996; Abramo et al. 2009; Postiglione and Jisun 2013; Marquina and Ferreiro 2015), our goal was to explore the "European research elite" through large-scale quantitative material. We sought to empirically test the expectations arising out of prior smaller-scale and single-nation research.

We explore both the intra-national differences in research productivity between this European research elite and the rest of research-involved academics (or "average" academics, as they are termed in Stephan and Levin 1992: 57–58 and Prpić 1996: 185), and cross-national differences and similarities among this European elite. Following prior research on the predictors of research productivity (especially Allison and Stewart 1974; Fox 1983; Stephan and Levin 1992; Ramsden 1994; Teodorescu 2000; Lee and Bozeman 2005; and recently Leisyte and Dee 2012; Shin and Cummings 2010; Drennan et al. 2013), our guiding question is as follows: How different are highly productive academics from "average" academics, how differently do they work and perceive their work, and which factors are positively correlated with high research performance?

This paper is structured as follows: Second section is entitled "Analytical framework" (with subsections on theories of academic productivity, and a review of the literature on highly productive academics, both qualitative and quantitative). Section three is focused on "Data and methods." The "Research findings" are presented in two subsections of fourth section: The first subsection reports the results from a bivariate analysis (on the two major correlates of high research productivity: high research time investment and high research role orientation), and the second subsection reports the results from a logistic regression analysis. Finally, there is fifth section "Discussion" and sixth section "Conclusions".

# Analytical framework

## Theories of research productivity

Research productivity has been an important scholarly topic for a long time (for some original formulations, see Crane 1965; de Solla Price 1963; Merton 1968; Cole and Cole 1973). The literature has identified a number of individual and institutional factors that influence research productivity, including the size of the department, disciplinary norms, reward and prestige systems, and individual-level psychological constructs such as a desire for the intrinsic rewards of puzzle-solving (see Leisyte and Dee 2012; Stephan and Levin 1992; Ramsden 1994; Teodorescu 2000, Kwiek 2015b). Faculty orientation toward research is generally believed to predict higher research productivity; as are the time spent on research, being a male academic, faculty collaboration, faculty academic training, years passed since PhD, as well as a cooperative climate and support at the institutional level (Porter and Umbach 2001; Katz and Martin 1997; Smeby and Try 2005; Lee and Bozeman 2005). The extreme differences in individual research productivity can be explained by a number of theories.

First, the "sacred spark" theory presented by Cole and Cole (1973) simply says "that there are substantial, predetermined differences among scientists in their ability and motivation to do creative scientific research" (Allison and Stewart 1974: 596). Highly productive scholars are "motivated by an inner drive to do science and by a sheer love of the work" (Cole and Cole 1973: 62). Productive scientists are a strongly motivated group of researchers, and they have the stamina, "or the capacity to work hard and persists in the pursuit of long-range goals" (Fox 1983: 287).

Second, the "accumulative advantage" theory developed by Robert K. Merton (1968) claims that productive scientists are likely to be even more productive in the future, while the productivity of those with low performance will be even lower. The accumulative advantage theory is related to the reinforcement theory formulated by Cole and Cole (1973: 114) which in its simplest formulation states that "scientists who are rewarded are productive, and scientists who are not rewarded become less productive." As Gaston (1978: 144) points out, reinforcement deals with *why* scientists continue in research activities, and accumulative advantage deals with *how* some scientists are able to obtain resources for research that in turn leads to successful research and publication. Several studies (Allison and Stewart 1974; Allison et al. 1982) support the cumulative advantage hypothesis, without discrediting the sacred spark hypothesis.

Finally, according to the "utility maximizing theory," all researchers choose to reduce their research efforts over time because they think other tasks may be more advantageous. As Kyvik (1990: 40) comments, "eminent researchers may have few incentives to write a new article or book, as that will not really improve the high professional reputation that they already have." And Stephan and Levin (1992: 35) in discussing age and productivity argue that "with each additional year, the rewards for doing research decline." These three major theories of research productivity are complementary rather than competing: To varying degrees, they are all applicable to the academic profession.

### Highly productive academics: review of the literature

We distinguish two different approaches in the research literature for exploring individual-level high research productivity. The first approach was to explore it through qualitative material: First, rankings of highly productive academics are created, and then, academics in the top ranks are interviewed, with a general research question of "how can they be so productive?" (Mayrath 2008: 42). Various "keys to productivity" (Kiewra and Creswell 2000: 155) or "guidelines for publishing" (Kiewra 1994) are drawn. Qualitative studies present a large number of useful tips and refer to some striking individual examples. However, conversation-based qualitative explorations though fascinating are somehow under-theorized. The second approach, in contrast, was to explore high research productivity through quantitative material: surveys in which (academic) behavioral and attitudinal data are combined with publication data. In this paper, we shall use the second, quantitative approach.

Faculty research productivity has been thoroughly explored, mostly in single-nation contexts: especially the USA, the United Kingdom, and Australia (Cole and Cole 1973;

Allison and Stewart 1974; Fox 1983; Ramsden 1994), as well as South Korea (Shin and Cummings 2010), but rarely in cross-national contexts (exceptions include Teodorescu 2000; Drennan et al. 2013; Postiglione and Jisun 2013). While most productivity studies focused on faculty from selected academic fields, especially from the natural sciences, our study uses national samples and refers to all academic fields grouped into five large clusters.

International comparative studies in higher education have not generally explored a specific class of highly productive academics; however, they have been mentioned in passing in several single-nation academic profession studies (Crane 1965; Cole and Cole 1973; Allison 1980), but they were not researched in more detail in these studies. Exceptions include a discussion of American "big producers" in *Little Science, Big Science* by Derek J. de Solla Price (1963); a study of "star scientists" in the context of gender differences in research productivity in Italy in Abramo et al. (2009); and studies in the productivity of Croatian "eminent scientists" in Prpić (1996). Abramo et al. (2009: 143) conclude that a star scientist "is typically a male full professor." They argue that "to obtain levels of scientific production such as those of a star scientist, the time and energy required for research activities are notably superior to the average, and imply an overwhelming dedication to work" (Abramo et al. 2009: 154). However, as their work is based on bibliometric data, the authors are unable to go beyond gender, academic rank, institutional type, and academic discipline in their exploration of a "star scientist profile." Prpić (1996) compared the scientific productivity of "eminent" and "average" scientists. Her research assumptions were that the patterns of predictors for the publication productivity of eminent scientists would be different from those of "average" scientists because in the elite group, "homogeneity is larger and variability is smaller than in the entire research population" (Prpić 1996: 199).

Recently, Postiglione and Jisun (2013) studied "top tier researchers" in four Asian countries, seeking commonalities shared by them based on the CAP ("Changing Academic Profession") survey. They studied 10 % of the most and least productive academics through descriptive statistics. They found that highly productive academics emphasize basic/theoretical research and social responsibility in science more often than the rest of academics, and spend more time on research than on teaching (Postiglione and Jisun 2013: 171–177). Also, Marquina and Ferreiro (2015) studied a specifically constructed "elite group" of academics in six "emergent" countries, based on the same global academic profession survey. They compared "elite groups" with the "rest" of academics. Their class of "elite groups" does not refer directly to research productivity, there are important parallels, though: academics in "elite groups" are more internationalized in teaching, research, and publishing, and they spend more time on research and are more research-oriented (Marquina and Ferreiro 2015: 191). The major theories of research productivity as well as studies on highly productive academics, rare as they have been, provide conceptual underpinning for the present study.

## Data and methods

### Data

We explore research productivity defined here, following Daniel Teodorescu (2000: 206), as the "self-reported number of journal articles and chapters in academic books that the respondent had published in the 3 years prior to the survey." The data come from the countries involved in both the CAP and EUROAC projects, with national datasets subsequently cleaned, weighted, and merged into a single European dataset.[1] We base our study empirically on the single most important cross-national source of data on academic views, attitudes, perceptions, and behaviors in Europe available today, with all its inherent limitations for comparative research. The quality of the dataset is high (Teichler et al. 2013: 35; Teichler and Höhle 2013: 9) as well as being well suited for our research purposes.

The survey questionnaire was sent out to the CAP countries in 2007 and to the EUROAC countries in most cases, including Poland, in 2010. The total number of returned surveys was 17,211 and included between 1000 and 1700 returned surveys from all countries studied except for Poland where it was higher, as shown in Table 1 in the "Electronic Supplementary Material," or ESM. Overall, the response rate differed from over 30 % in Norway, Italy, and Germany; to 20–30 % in the Netherlands, Finland, and Ireland; to about 15 % in the United Kingdom; 11 % in Poland; and 10 % or less in Austria, Switzerland, and Portugal. There are no indications that the pool of respondents differs from the pool of non-respondents, and consequently, the "non-response bias" (Stoop 2012: 122) did not seem to occur in any of the countries. Overall, simple random sampling, systematic sampling, and stratified random sampling methods were used, depending on the country (national-level sampling techniques are described for the CAP European countries in RIHE 2008: 89–178, and for the EUROAC countries in Teichler and Höhle 2013: 6–9). The proportion of faculty by academic field cluster is given in Table 8 of the ESM, and composite country indexes for the research productivity of full-time academics employed in the university sector in Table 9 in the ESM).

## Methods

We divided the sample of all European academics into two complementary subsamples: academics reporting research involvement and those not reporting this. Then, the subsample of research-involved academics was divided into two further subgroups: The first was "research top performers" (henceforth referred to as "top performers"), identified as academics ranked among the top 10 % (cutoff points permitting) of academics with the highest research performance in each of the 11 national systems (separately) and in all the five major research field clusters (also separately). The second subgroup was that of the remaining 90 % of academics involved in research. The distribution of the sample population is shown by country in Table 1.

Top performers, as defined in this paper, provide substance to European research production: Without them, it would be halved. Because, on average, consistently across all European systems studied, slightly less than half (45.9 %) of all academic research production as measured by journal articles comes from about 10 % of the most highly productive academics. In four systems, the share is close to, or exceeds, 50 % (Austria, Finland, Poland, and Portugal, see Table 2; the upper 5 % of highly productive academics show a similar pattern: They produced on average 33 % of all journal articles).

Treating the consistent patterns of productivity distribution found above as a starting point in this research, to begin with we shall discuss top performers through a bivariate analysis of the working time distribution and the teaching/research role orientation.

---

[1] We worked on the final dataset dated June 17, 2011 created by René Kooij and Florian Löwenstein from the International Centre of Higher Education and Research—INCHER-Kassel.

**Table 1** Distribution of the sample population, by country

|  | All | Research-involved (N) | % Research-involved | Top performers | % Top performers |
|---|---|---|---|---|---|
| Austria | 1492 | 1297 | 86.9 | 146 | 11.3 |
| Finland | 1374 | 1063 | 77.4 | 126 | 11.9 |
| Germany | 1215 | 1007 | 82.9 | 110 | 10.9 |
| Ireland | 1126 | 865 | 76.8 | 101 | 11.7 |
| Italy | 1711 | 1674 | 97.8 | 191 | 11.4 |
| The Netherlands | 1209 | 536 | 44.3 | 61 | 11.4 |
| Norway | 986 | 876 | 88.8 | 106 | 12.1 |
| Poland | 3704 | 3659 | 98.8 | 411 | 11.2 |
| Portugal | 1513 | 944 | 62.4 | 104 | 11.0 |
| Switzerland | 1414 | 1210 | 85.6 | 138 | 11.4 |
| United Kingdom | 1467 | 777 | 53.0 | 89 | 11.5 |
| Total | 17,211 | 13,908 | 80.8 | 1583 | 11.4 |

**Table 2** Journal articles produced in the three-year reference period, by top performers and by the rest of academics, by country

|  | By top performers | By the rest | Total | % By top performers |
|---|---|---|---|---|
| Austria | 3330 | 1206 | 4536 | 73.4 |
| Finland | 2445 | 2435 | 4880 | 50.1 |
| Germany | 2702 | 3506 | 6208 | 43.5 |
| Ireland | 2419 | 2684 | 5103 | 47.4 |
| Italy | 5096 | 10,162 | 15,259 | 33.4 |
| The Netherlands | 1513 | 1647 | 3160 | 47.9 |
| Norway | 1902 | 2340 | 4243 | 44.8 |
| Poland | 6767 | 6831 | 13,599 | 49.8 |
| Portugal | 1992 | 1952 | 3945 | 50.5 |
| Switzerland | 2798 | 3304 | 6102 | 45.9 |
| United Kingdom | 1740 | 2475 | 4215 | 41.3 |
| Total | 32,706 | 38,543 | 71,248 | 45.9 |

Although bivariate analyses are limited insofar as they do not control for other important factors that might affect research productivity (Teodorescu 2000: 203), the two selected variables emerge as key in most qualitative and quantitative productivity studies. However, a study of multidimensional relationships requires a model approach, and therefore, odds ratio estimates by logistic regression for belonging to European research elites will be presented, following inferential analyses. Inferential analyses and logistic regression analyses are viewed here as complementary: Both approaches are useful for our research purposes.

More specifically, in the section on the working time distribution, an independent two-sample $t$ test is used. When the variance in the compared populations is equal (Levene's test of homogeneity of variance is used), then Student's $t$ test is used; otherwise, Welch's two-sample $t$ test is used. The test statistic has a $t$ distribution. Consistent with previous research on the "working time distribution" in academia (Bentley and Kyvik 2013), we focus here on annualized weekly hours in both teaching periods of the academic year and in non-teaching periods: 60 % for the former and 40 % for the latter time as a good approximation for the majority of the European systems studied. In the section on teaching/research role orientation, in order to compare fractions, a two-proportion $z$ test is used. The test statistic has a standardized normal distribution. All tests are conducted with a significance level of $\alpha = 0.05$.

## Limitations

First, all the publication data are self-reported, and the differences in reporting them can be between nations, academic disciplines, and genders: Consequently, to different degrees, respondents "may present an untrue picture to the researcher, for example answering what they would like a situation to be rather than what the actual situation is" (Cohen et al. 2011: 404). Although self-reported publication data are not perfect, they do not seem to be subject to any systemic error or systemic bias. Second, due to the anonymization of the collected data, we were unable to study any differences between top performers from institutions of lower academic standing and those from the most prestigious ones. The third limitation comes from a tacit assumption that the major concepts used in the survey instrument in all systems have a somehow similar definition. Another limitation is inherent to the structure of the dataset used: In regression analysis, no distinction can be made between single-authored and multiple-authored publications and between national and international publications (except through various proxies).

## Research findings

### Bivariate analysis: academic behaviors and attitudes

The first question in this section is whether the working habits of top performers are different from those of the remaining 90 % of research-involved academics. The second question is whether top performers are more research-oriented (both consistent with the research literature on research productivity, see especially Fox 1992; Bentley and Kyvik 2013; Shin and Cummings 2010).

*Academic behaviors: working time distribution*

We explore here the five dimensions of academic work which were captured by the CAP/EUROAC datasets: teaching, research, service, administration, and "other" academic activities. The mean for the annualized total working time differential between top performers and the rest of academics is 6.2 h, ranging from 2.2 h in Italy to 9.4 h in Norway and 10.2 h in Germany (see the details by type of academic activity and by country in Table 12 in the ESM). In other words, for example, German top performers, when compared with the rest of (research-involved, as in the whole paper) German academics, spend on average an additional 66.3 full working days in academia per year (10.2 h times 52 weeks divided by 8 h per day), and Norwegian top performers spend on average an additional 61.1 full working days.

We know from previous research productivity studies that longer working hours, and especially research hours, substantially contribute to high productivity: Our study shows (with powerful results: *p* value <0.001) what exactly "longer hours" mean for the upper 10 % of highly productive academics, and shows it from a comparative cross-national perspective. A ticket to enter the class of national top performers differs from country to country, though, as systems are not equally competitive: In more competitive systems, top performers work much longer hours than in less competitive systems when compared to average academics.

We are interested in the differences in the means of total working hours, and especially the means of research hours, between the two subpopulations in each country and the significance of the results (Table 3). Our results are based on two-sided tests assuming equal differences in arithmetic means with a significance level $\alpha = 0.05$. For each pair with a mean difference significantly different from zero, the symbol of the larger category ("Top" for top performers or "Rest" for the rest of academics) appears in the column. Tests are adjusted for all pairwise comparisons within a row for each innermost subtable using the Bonferroni correction. The *t* tests for the equality of two arithmetic means (Top vs. Rest) were performed for each country for each of the five types of academic activities studied.

As clearly shown in Table 3, longer research hours for top performers are statistically significant for a pool of seven countries ("Top" symbols in the line of "research," the only exception being Switzerland). But also for a pool of seven countries, longer administration hours for top performers are statistically significant ("Top" symbols in the line of "administration"). The same applies to service hours (three countries) and hours spent on "other" academic activities (four countries). Not surprisingly, the same also applies to total working hours in all the countries studied. In three countries (Austria, Norway, and Switzerland), their longer teaching hours are also statistically significant. Further details are given in Table 12 in the ESM: In the column "group with significantly larger mean," top performers appear in respect of almost all countries, that is, they work longer hours in almost all the categories studied. There is a standard working pattern for top performers in most of the countries studied. "Science takes time," and much more scientific production takes much more time. Top performers work (much) longer hours. Their longer total working time is statistically significant for all countries.

*Academic attitudes: teaching and research orientation*

The results of the *z* test for the equality of fractions performed for all countries (Table 4) are based on two-sided tests with a significance level of $\alpha = 0.05$. Tests are adjusted for all pairwise comparisons within a row for each innermost subtable using the Bonferroni correction.

The *z* tests for the equality of fractions (Top vs. Rest) were performed for each country for each of the four categories of teaching and research orientation. Correspondingly, as before, for each pair with a fraction difference significantly different from zero, the symbol for the larger category ("Top" for research top performers or "Rest" for the rest of academics) appears in the column.

As clearly shown in Table 4, the research role orientation (answer 3) among top performers is statistically significant in a pool of eight countries ("Top" symbols in the line for "in both, but leaning toward research", with no exceptions). Additionally, in a pool of five countries, the strong research role orientation (answer 4) for top performers is also statistically significant, again with no exceptions. The division in role orientation between

**Table 3** Results of $t$ tests for the equality of means, top performers (Top) versus the rest of academics (Rest), all countries

|  | AT | FI | DE | IE | IT | NL | NO | PL | PT | CH | UK |
|---|---|---|---|---|---|---|---|---|---|---|---|
| Teaching | Top |  |  | Rest | Rest |  | Top | Rest |  | Top |  |
| Research |  |  | Top | Top | Top | Top |  | Top | Top | Rest | Top |
| Service |  |  |  |  | Top |  | Top |  |  | Top |  |
| Administration | Top | Top | Top |  |  | Top | Top | Top |  | Top |  |
| Other |  |  |  | Top | Top |  | Top |  |  | Top |  |
| Total | Top | Top | Top | Top | Top | Top | Top | Top | Top | Top | Top |

How long do you spend on various academic activities, only full-time academics in universities involved in research (mean per year, 60 % when classes are in session and 40 % when classes are not in session)

**Table 4** Results of $z$ tests for the equality of fractions, all countries

|  | AT | FI | DE | IE | IT | NL | NO | PL | PT | CH | UK |
|---|---|---|---|---|---|---|---|---|---|---|---|
| Primarily in teaching |  | Rest | Rest | Rest | –[a] | –[a] | –[a] | Rest | Rest |  | –[a] |
| In both, but leaning toward teaching |  | Rest |  | Rest | Rest | Rest | Rest | Rest | Rest |  | Rest |
| In both, but leaning toward research |  | Top | Top | Top | Top | Top |  | Top | Top | Top |  |
| Primarily in research |  |  |  |  | Top | Top |  | Top | Top |  | Top |

Preferences for teaching/research (Question B2: "Regarding your own preferences, do your interests lie primarily in teaching or in research?"), research top performers (Top) versus the rest of academics (Rest)

[a] This category is not used in comparisons because its column proportion is equal to zero or one

top performers and the rest of academics is clear (and all differences are statistically significant): In all the systems studied, top performers are more research-oriented than the rest of academics (see the column "group with a larger mean" and answers 3 and 4 in Table 13 in the ESM). Being interested "primarily in teaching" virtually excludes such European academics from the class of research top performers: Their share attains a maximum of 2 % in Ireland but in the majority of them it is 0 %. In addition, being interested "in both, but leaning toward teaching" again almost excludes such European academics from the same class: Their share is about 3 % in the United Kingdom and 5–9 % in the other countries with only two exceptions: Poland (17.4 %) and Portugal (21.7 %) where it is substantially higher (both are teaching-oriented systems, though, Kwiek 2012, 2013a). Our results show that a research role orientation is a powerful indicator of belonging to the class of the European research elite: Being research-oriented is virtually a must for European academics and being teaching-oriented virtually excludes them from this class.

However, a study of multidimensional relationships requires a model approach with a number of dependent variables, including research hours and research orientation, among several others.

### Logistic regression analysis

We have developed an analytical model to study research productivity based on the research literature, especially Fox (1992: 295–297), Ramsden (1994: 211–212), and Teodorescu (2000: 207). Following Ramsden (1994), we have assumed that "any sensible explanation of research output must take into account personal (individual) and structural (environmental) factors, and preferably also the interaction between them." Independent variables are grouped as "individual" and "institutional" characteristics in eight clusters (see Table 5).

There are two questions related to the overall research approach taken. The first question is why estimating a regression model for each of the 11 countries rather than pooling the sample and control for country. The argument for the choice of 10 % top performers per country (and per major academic field cluster) is that the approach of selecting merely the upper 10 % of academics, regardless of the country, does not fit the purpose of highlighting cross-national differences among top performers. The factors

Table 5 Faculty research productivity: variables in the model

| Individual variables | Institutional variables |
| --- | --- |
| *Personal/demographics* | *Institutional policies* |
| Female (F1) | Strong performance orientation (E4) |
| Mean age (F2) | Research considered in personnel decisions (E6) |
| Full time (A7) | *Institutional support* |
| Professor (A10) | Availability of research funds (B3) |
| *Socialization* | Supportive attitude of administration (E4) |
| Intensive faculty guidance (A3) | |
| Research projects with faculty (A3) | |
| *Internationalization and collaboration* | |
| Collaborating internationally (D1) | |
| Collaborating domestically (D1) | |
| Publishing in a foreign country (D5) | |
| Research int'l in scope or orientation (D2) | |
| *Academic behaviors* | |
| Annualized mean research hours (60 % in session and 40 % not in session) (B1) | |
| *Academic attitudes and role orientation* | |
| Research-oriented (only answer 4) (B2) | |
| Scholarship is original research (B5) | |
| Basic/theoretical research (D2) | |
| *Overall research engagement* | |
| National/int'l. committees/boards/bodies (A13) | |
| A peer reviewer (A13) | |
| Editor of journals/book series (A13) | |

Survey question numbers in parentheses

Table 6 Odds ratio estimates by logistic regression for being in the top 10 % in research productivity, all countries

| | Austria | Finland | Germany | Ireland | Italy | The Netherlands | Norway | Poland | Portugal | Switzerland | United Kingdom |
|---|---|---|---|---|---|---|---|---|---|---|---|
| Nagelkerke's $R^2$ | 0.129 | 0.301 | 0.258 | 0.302 | 0.157 | 0.248 | 0.32 | 0.246 | 0.538 | 0.312 | 0.402 |
| Percentage of cases predicted correctly | 81.9 | 78.6 | 83.7 | 86.8 | 87.2 | 81.1 | 80.1 | 80.3 | 85.3 | 77.4 | 87.5 |
| *Individual predictors* | | | | | | | | | | | |
| Personal/demographics | | | | | | | | | | | |
| Female | | | | | 0.566* | | | | | | 0.379** |
| Age | | | | 1.035* | | | | 0.963*** | 1.105*** | 1.034* | |
| Full time | | | | | | | | | 0.07**[a] | | |
| Professor | | 3.381*** | 3.242*** | 3.682*** | | 2.681* | 3.661*** | 1.799* | | | |
| Socialization | | | | | | | | | | | |
| Intensive faculty guidance | | | | – | 0.526* | | | 0.707* | | | |
| Research projects with faculty | | | | – | | | | | 7.761***[a] | | |
| Internationalization and collaboration | | | | | | | | | | | |
| Collaborating internationally | | | 2.135* | 2.485** | 3.165* | 4.134** | | 1.699** | | | |
| Collaborating domestically | | 11.089**[a] | | – | | – | 5.461** | | | | |
| Publishing in a foreign country | | | | 3.524* | | | | 2.185*** | 7.436*[a] | 3.377** | 4.237** |
| Research int'l. in scope or orientation | 2.106** | | 0.49** | | 2.325* | | | | 4.589** | | |
| Academic behaviors | | | | | | | | | | | |
| Annualized mean weekly research hours (60 % in session, 40 % not in session) | | | 1.029** | | | | 1.026* | | | | 1.037** |

Table 6 continued

| | Austria | Finland | Germany | Ireland | Italy | The Netherlands | Norway | Poland | Portugal | Switzerland | United Kingdom |
|---|---|---|---|---|---|---|---|---|---|---|---|
| *Academic attitudes and role orientation* | | | | | | | | | | | |
| Research-oriented | | | | 3.141*** | | | | 1.51* | | | |
| Scholarship is original research | | | | 0.549* | | | | | | – | |
| Basic/theoretical research | | | | 2.231** | 1.862** | | | | 3.183* | | |
| Overall research engagement | | | | | | | | | | | |
| Nat./int'l. committees/boards/bodies | 2.474*** | | | | | | | 1.833*** | | 2.887** | 2.263* |
| A peer reviewer | 1.652* | | | 9.038***[a] | | | 2.741* | | | | 8.529*[a] |
| Editor of journals/book series | | | | | | | | 2.815*** | | 3.4*** | 2.372** |
| *Institutional predictors* | | | | | | | | | | | |
| Institutional policies | | | – | | | | | | | | |
| Strong performance orientation | | | | | | | | | | 2.009** | |
| Research consid. in HR decisions | | | | | | | | | 0.177**[a] | – | 2.216* |
| Institutional support | | | | | | | | | | | |
| Availability of research funds | | | | | | | | | | | |
| Supportive attitude of administration | | | | | | | | | | | |

Results that are not statistically significant are not shown in the Table

"–" No usable data available (question was not asked).

*** $p < 0.001$; ** $p < 0.01$; * $p < 0.05$.

[a] These odds ratios need to be treated with caution

important in predicting high research productivity in some countries might be irrelevant in other countries. However, we have also developed a single model controlling for country fixed effects, and the two models will be compared briefly in the "Discussion" section. The second question is why the regression model is not controlled for academic discipline as a potentially important source of variation: Unfortunately, the number of observations per discipline was too small in many cases.

In this multivariate analysis, we have dichotomized all category variables through a recoding procedure. We started with 42 personal and institutional characteristics, grouped into eight clusters (see Table 10 in the ESM). We then conducted Pearson Rho's correlation tests to find significantly correlated predictors of the dependent variable. The predictors were entered into a four-stage logistic regression model (as in Cummings and Finkelstein 2012). Multicollinearity was tested using an inverse correlation matrix, and no independent variables strongly correlated with others were found. The predictive power of the fourth model (as measured by Nagelkerke's $R^2$) was the highest for Portugal (0.54), the United Kingdom (0.40), Norway, Ireland, Switzerland, and Finland (about 0.30–0.32). The total average variance demonstrated for the 11 countries studied is about 32 %. The predictive power of the models of research productivity estimated by other researchers is not substantially higher (the average variance in Drennan et al. 2013: 129 is about 30 %, and about 30 % for 10 globally studied countries in Teodorescu 2000: 212). A model fit defined via the percentage of cases predicted correctly is generally in the 80–90 % range. In Table 6, we present the results of the final, fourth model.

### Statistically significant individual variables

The collection of individual variables emerges as more important than the collection of institutional variables, both in terms of the frequency of occurrence and the size of regression coefficients.

In the first block of individual predictors ("personal/demographics"), being a female academic entered the equation in two countries only: It is a strong predictor of *not* becoming a top performer in Italy, where the odds ratio value indicates that female academics are about half as likely as male academics to be a top performer, and in the United Kingdom, where they are only about one-third as likely. But in all other countries, being a male academic is not a predictor of becoming a top performer. While the finding for Italy is consistent with the analysis of Italian "star scientists" in Abramo et al. (2009), overall, our findings are clearly different from the findings from linear regression analyses in which being a female academic has traditionally been negatively correlated with research productivity.

While in most single-nation and cross-national studies age is not a statistically significant variable, our model shows that "age" is a powerful predictor of high research performance in four countries. A one-unit increase (i.e., 1 year) in Ireland and Switzerland increases the odds of becoming a top performer by about 3.5 % on average (*ceteris paribus*) and in Portugal by 10.5 %.

Finally, being a "professor" (or academic seniority) emerged as the single most important variable in the model, with statistical significance in six countries. In four of them (Finland, Germany, Ireland, and Norway), being faculty at senior ranks increases the odds of becoming a top performer more than three times, in the Netherlands slightly less than three times, and in Poland almost twice (see Kwiek and Antonowicz 2015). This finding confirms the conclusions from previous productivity studies, although certainly academics in European higher education are more likely to be promoted to higher ranks if they are highly productive. Productivity affects being a professor and the relationship may be "reciprocal" (Teodorescu 2000: 214). But as Ramsden (1994: 223) argued, "identifying correlates of high productivity does not mean that we have identified causal relations."

In the block of individual predictors, "socialization," to great surprise, especially in the context of the US literature, both variables are either statistically insignificant or, as in two countries (Poland and Italy), they actually *decrease* the odds of becoming top performers. It could be that in an "academic oligarchy" types of systems (Kwiek 2013b, 2015a), doctoral students receive faculty guidance more by working for senior faculty, possibly as a cheap academic labor force, rather than independently working with them (and later productivity is substantially influenced by the early recognition of research work). The block of "internationalization and collaboration" emerges as the single most important grouping in predicting high research productivity: Each of the four variables at least doubles the odds of becoming a top performer. The four variables are as follows: "collaborating internationally," "collaborating domestically," "publishing in a foreign country," and "research international in scope or orientation." These variables enter the equation in all countries except one (Finland).

Domestic collaboration, as opposed to international collaboration, does not influence high research productivity in any country except for the United Kingdom. "Publishing in a foreign country" emerged as a powerful predictor in four smaller higher education systems: Ireland, Poland, Switzerland, and Norway, as with small academic markets it makes it more necessary for prolific academics to publish internationally. Also, "research international in scope or orientation" increases the odds in three countries. The atypical case of Germany where this variable actually decreases by half the odds of being a top performer could be explained by the large size of the national publishing academic market.

In the block of "academic behaviors," contrary to previous research conclusions from linear regression models (most recently in Cummings and Finkelstein 2012: 58; Shin and Cummings 2010: 590; Drennan et al. 2013: 127), annualized mean weekly research hours emerged as determinative predictors only in three countries (Germany, Norway, and the United Kingdom): A unit increase of 1 h (in annualized research hours per week) increases the odds of being a top performer by a 2.6–3.7 % on average (*ceteris paribus*). In all the other countries, a high research time investment is not a determinative predictor of becoming a top performer.

Again, in the block of "academic attitudes and role orientation," contrary to the findings from previous linear regression models, research orientation emerged as a powerful predictor of research productivity in only two countries, with Exp(B) = 3.141 for Ireland and Exp(B) = 1.51 for Poland. In all other countries, it was not a determinative predictor.

Surprisingly, while in simple descriptive statistics (both here and in Postiglione and Jisun 2013) and in inferential analyses presented above, both long research hours and high research orientation emerge as important characteristics of top performers, following the almost universal findings in the research productivity literature, here, a multidimensional model approach supports these findings in selected countries only.

### Statistically significant institutional variables

The importance of variables differs from country to country, but the overall determinative power of individual-level predictors is much stronger than those of institutional-level predictors, consistent with previous research on productivity (Ramsden 1994: 220; Shin and Cummings 2010: 588; Teodorescu 2000: 212; Cummings and Finkelstein 2012: 59). As Drennan et al. (2013: 128) concluded, "institutional factors were found to have very little impact on research productivity." This finding is also consistent with the conclusion about the American professoriate that "intrinsic motivations" rather than "institutional incentive structures" (Finkelstein 1984: 97–98, Teodorescu 2000: 217) stimulate research productivity. In general, the institutional-level predictors are statistically significant in only two cases in two countries (Switzerland and the United Kingdom). Surprisingly in the context of previous research (Fox 1983), two institutional predictors are not statistically significant in any of the countries studied: "availability of research funds" and "supportive attitude of administration." This might mean that, generally, neither institutional policies nor institutional support substantially matter in becoming a top performer.

Interestingly, while the conclusions from linear regression models indicate that institutional-level predictors of research productivity are weak, in our logistic regression model the conclusions indicate that they are actually statistically insignificant. In particular, research funds and academic climate (good academic–administration relationships) do not enter the equations in any country in the model (on collegiality across Europe, see Kwiek 2015a). Also, the strong performance orientation of institutions is statistically insignificant in all countries except Switzerland.

## Discussion

The findings from statistical inference show two clear cross-national patterns applicable to top performers: longer working hours (in all time categories) and higher research orientation. In only three countries do the rest of academics actually spending more time than top performers on any of the studied activities: This is teaching in Ireland, Italy, and Poland. The results from these three countries provide strong support for a thesis about an antagonistic or competitive relationship between teaching and research (as argued by Fox 1992 who discussed "mutuality" and "competition" between teaching and research), at statistically significant levels: While highly productive academics in these countries spend more time on research, the rest of academics spend more time on teaching. In these countries, as Fox (1992: 303) argued, teaching and research "are at some odds with each other." Top performers work (much) longer hours, and their longer total working time is statistically significant for all countries. From a statistical inference approach, top performers are also more research-oriented than the rest of academics. The most salient difference between the two subpopulations can be seen in three structurally similar systems having a similar teaching/research time distribution: In Ireland, Poland, and Portugal, only about half of the "rest" of academics is research-oriented. They are nominally involved in research but not research-oriented in their self-declared role preferences. In general, the distribution of research role orientation is almost universal across all the countries studied. Consequently, highly productive academics are almost universally more intra-nationally different from "average" academics, and almost universally more similar to top performers in other countries.

There are important differences in the conclusions from linear regression models detailed in previous studies, and the conclusions derived via a multiple regression model from predictors of belonging to a distinctive group of the European research elite as defined in this paper. The internationalization of research, national and international research collaboration, international publishing, academic seniority, as well as high levels of overall research engagement emerge as powerful correlates of high research productivity (on productivity of "internationals" contrasted with "locals" across 11 European systems, see Kwiek 2015b). Also, in both cases, the overall determinative power of individual-level predictors is stronger than that of institutional-level predictors (as in Ramsden 1994: 223; Shin and Cummings 2010: 586; Cummings and Finkelstein 2012: 58).

While both in the first approach, through descriptive statistics, and in the second approach, through $t$ test and $z$ test analyses research hours and research orientation strongly characterize top performers, a multidimensional model approach through regression analysis, surprisingly, supports these findings in selected countries only. From among individual variables, both age and academic seniority are important predictors of high research productivity. However, neither annualized research hours, nor research orientation (traditionally, the two most important predictors of research productivity) emerged as powerful predictors of high research productivity in more than three and two countries, respectively. This is perhaps the most perplexing result of our research: While in inferential analyses these are critical variables in all the systems studied, in multidimensional analyses, their role is considerably smaller than expected. The specific case of working time distribution and research role orientation clearly shows that a combination of several approaches is more fruitful than a reliance on any of them separately.

There is also a tension between the conclusions drawn from our 11 multiple regression models and the single model controlling for country fixed effects (see Table 13 in the ESM). The difference is in focus: highly productive academics being explored as nested in the context of national systems or explored independent of the context (the sample distribution is given in Table 14 in the ESM). While in the first model, in the block of personal/demographic variables, both age and gender entered the equation in several countries, in the single model for European academics both were statistically insignificant. In both models, higher mean weekly research hours increase the odds [Exp(B) = 1.026–1.037 and Exp(B) = 1.017, respectively]. However, self-declared research role orientation in the first model is statistically significant in only two countries, and in the second, single model, it is statistically insignificant.

The differences in conclusions from our two different logistic regression models (with top performers differently defined, in Europe as a whole or separately in European systems) are smaller than expected: In the context of previous single-nation studies, the insignificance of both age and gender in the single model comes as a surprise. The emergence of academic seniority as a predictor of high research productivity in the single model is consistent with previous studies, but the statistical significance of the research role orientation in only two countries in the first model and its insignificance in the single model come as a surprise. This may imply that there is a growing tension between self-declared research role orientation and research productivity in Europe. While European academics increasingly view themselves as research-oriented, research orientation emerges as a much less statistically significant predictor of becoming a top performer than expected from previous research (see Kwiek 2015b). In contrast, research time investments emerge as significant predictors in both the first model (in three countries) and in the single model.

The overall relative insignificance of institutional predictors (in both models) in the case of highly productive academics may provide further support for the "sacred spark" theory of productivity (Cole and Cole 1973): Regardless of administrative and financial institutional settings, some faculty—and they may be our "research top performers"—will always show greater inner drive toward research than others. Also, Bentley and Kyvik (2013) in their global study of 13 countries found more support for this theory than for the competing "utility maximization theory." The "accumulative advantage" theory (combined with "reinforcement theory") found only partial support in the study: Age is not a significant predictor in most systems studied, and academic seniority (or professorships), although a significant predictor in most systems, is reciprocally linked to productivity.

## Conclusions

The role of highly productive academics in knowledge production across all 11 European systems studied is pivotal: Without these 10 % of academics, national academic outputs would be halved. We have presented an international comparative study based on solid quantitative material rather than the single-nation studies that dominate previous research. In contrast to bibliometric studies of research productivity, we focused on academic attitudes, behaviors, and perceptions as the predictors of becoming research top performers. Our study provides a large-scale and cross-national corroboration of the systematic inequality in knowledge production, suggested for the first time by Lotka (1929) and de Solla Price (1963). What we may term the "10/50 rule" holds strongly across Europe (with the upper 10 % of academics producing about 50 % of all publications, see also Kwiek 2015c).

The European research elite is a highly homogeneous group of academics whose high research performance is driven by structurally similar factors which cannot be easily replicated through policy measures. The variables increasing the odds of entering this class are individual rather than institutional. They come, they work according to similar working patterns, and they share similar academic attitudes. They are similar from a European cross-national perspective, and they substantially differ intra-nationally from their lower-performing colleagues. They are a universal academic species, and they share roughly the same burden of academic production across Europe (see Kwiek 2015c).

There are important differences in those conclusions from linear regression models with the correlates of research productivity detailed in previous studies and the conclusions from a multiple regression model with predictors of belonging to the European research elite. Our study shows the gender of academics as a very weak predictor, their age as a powerful predictor, and academic seniority and internationalization as the most important predictors. Contrary to most previous findings based on linear regression models, both annualized mean weekly research hours and research role orientation only emerged as powerful predictors of becoming a research top performer in several countries. In line with most previous research, though, institutional-level predictors emerged as statistically insignificant.

The study also shows a considerable tension between the conclusions from inferential results and logistic regression results. Surprisingly, while in inferential analyses both long research hours and high research orientation emerge as critical characteristics of top performers, a multidimensional model approach supports these findings in selected countries only. While in inferential analyses these are crucial variables in all the systems studied, in multidimensional analyses, their role is small. We conclude, therefore, that a combination of several approaches provides a better empirical insight into the European research elite. It is hard to entirely disregard the finding that being research-oriented is virtually a must to enter to the class of research top performers in Europe and being teaching-oriented virtually excludes European academics from this class. This finding has strong policy implications, especially for hiring new academic staff.

Therefore, based on the combination of inferential and multiple regression findings, European research top performers emerge in this study as much more cosmopolitan (the power of internationalization in research), much more hardworking (the power of long overall working hours and long research hours), and much more research-oriented (the power of a single academic focus) than the rest of European academics, despite differentiated national contexts.

**Acknowledgments** The author gratefully acknowledges the support of the National Research Council (NCN, Grant DEC-2011/02/A/HS6/00183). The author wishes to thank Ulrich Teichler, the coordinator of the European Science Foundation EUROAC project, "Academic Profession in Europe: Responses to Societal Challenges," and Dr. Wojciech Roszka for his assistance. Finally, the author wishes to express his gratitude to the three anonymous reviewers for their patience and a very constructive approach.

# References

Abramo, G., D'Angelo, C., & Caprasecca, A. (2009). The contribution of star scientists to overall sex differences in research productivity. *Scientometrics, 81*(1), 137–156.

Allison, P. (1980). Inequality and scientific productivity. *Social Studies of Science, 10*, 163–179.

Allison, P., Scott Long, J., & Krauze, T. (1982). Cumulative advantage and inequality in science. *American Sociological Review, 47*, 615–625.

Allison, P., & Stewart, J. (1974). Productivity differences among scientists: Evidence for accumulative advantage. *American Sociological Review, 39*, 596–606.

Bentley, P. J., & Kyvik, S. (2013). Individual differences in faculty research time allocations across 13 countries. *Research in Higher Education, 54*, 329–348.

Cohen, L., Manion, L., & Morrison, K. (2011). *Research methods in education*. New York: Routledge.

Cole, J., & Cole, S. (1973). *Social stratification in science*. Chicago: The University of Chicago Press.

Crane, D. (1965). Scientists at major and minor universities: A study of productivity and recognition. *American Sociological Review, 30*, 699–714.

Cummings, W. K., & Finkelstein, M. J. (2012). *Scholars in the changing American academy: New contexts, new rules and new roles*. Dordrecht: Springer.

de Solla Price, D. J. (1963). *Little science, big science*. New York: Columbia University Press.

Drennan, J., Clarke, M., Hyde, A., & Politis, Y. (2013). The research function of the academic profession in Europe. In U. Teichler & E. A. Höhle (Eds.), *The work situation of the academic profession in Europe: Findings of a survey in twelve countries* (pp. 109–136). Dordrecht: Springer.

Finkelstein, M. J. (1984). *The American Academic Profession. A synthesis of social scientific inquiry since World War II*. Columbus: Ohio State University Press.

Fox, M. (1983). Publication productivity among scientists: A critical review. *Social Studies of Science, 13*, 285–305.

Fox, M. F. (1992). Research, teaching, and publication productivity: Mutuality versus competition in academia. *Sociology of Education, 65*, 293–305.

Gaston, J. (1978). *The reward system in British and American science*. New York: Wiley.

Katz, J. S., & Martin, B. R. (1997). What is research collaboration? *Research Policy, 26*(1), 1–18.

Kiewra, K. A. (1994). A slice of advice. *Educational Researcher, 23*(3), 31–33.

Kiewra, K., & Creswell, J. (2000). Conversations with three highly productive educational psychologists: Richard Anderson, Richard Mayer, and Michael Pressley. *Educational Psychology Review, 12*(1), 135–161.

Kwiek, M. (2012). Changing higher education policies: From the deinstitutionalization to the reinstitutionalization of the research mission in polish universities. *Science and Public Policy, 39*, 641–654.

Kwiek, M. (2013a). From system expansion to system contraction: Access to higher education in Poland. *Comparative Education Review, 57*(3), 553–576.

Kwiek, M. (2013b). *Knowledge production in European Universities. States, markets, and academic entrepreneurialism*. Frankfurt and New York: Peter Lang.

Kwiek, M. (2015a). The unfading power of collegiality? University governance in Poland in a European comparative and quantitative perspective. *International Journal of Educational Development, 43*, 77–89.

Kwiek, M. (2015b). The internationalization of research in Europe. A quantitative study of 11 national systems from a micro-level perspective". *Journal of Studies in International Education*, OnlineFirst: February 25, 2015. doi:10.1177/1028315315572898

Kwiek, M. (2015c, forthcoming). Inequality in Academic knowledge production. In E. Reale, & E. Primeri, (Eds.), *The role of research top performers across Europe. Universities in transition. Shifting institutional and organizational boundaries*. Rotterdam: Sense.

Kwiek, M., & Antonowicz, D. (2015). The changing paths in academic careers in European universities: Minor steps and major milestones. In T. Fumasoli, G. Goastellec, & B. M. Kehm (Eds.), *Academic work and careers in Europe: Trends, challenges, perspectives* (pp. 41–68). Dordrecht: Springer.

Kyvik, S. (1990). Age and scientific productivity. Differences between fields of learning. *Higher Education, 19*(1), 37–55.

Lee, S., & Bozeman, B. (2005). The impact of research collaboration on scientific productivity. *Social Studies of Science, 35*(5), 673–702.

Leisyte, L., & Dee, J. (2012). Understanding academic work in changing institutional environment. *Higher Education: Handbook of Theory and Research, 27*, 123–206.

Lotka, A. (1929). The frequency distribution of scientific productivity. *Journal of Washington Academy of Sciences, 16*, 317–323.

Marquina, M., & Ferreiro, M. (2015). The academic profession: The dynamics of emerging countries. In W. K. Cummings & U. Teichler (Eds.), *The relevance of academic work in comparative perspective* (pp. 179–192). Dordrecht: Springer.

Mayrath, M. (2008). Attributions of productive authors in educational psychology journals. *Educational Psychology Review, 20*, 41–56.

Merton, R. (1968). The Matthew effect in science. *Science, 159*(3810), 56–63.

Porter, S. R., & Umbach, P. D. (2001). Analyzing faculty workload and using multilevel modeling. *Research in Higher Education, 42*(2), 171–196.

Postiglione, G., & Jisun, H. (2013). World class university and Asia's top tier researchers. In Q. Wang, Y. Cheng, & N. C. Liu (Eds.), *Building world-class universities. Different approaches to a shared goal* (pp. 161–180). Sense: Rotterdam.

Prpić, K. (1996). Characteristics and determinants of eminent scientists' productivity. *Scientometrics, 36*(2), 185–206.

Ramsden, P. (1994). Describing and explaining research productivity. *Higher Education, 28*, 207–226.

RIHE. (2008). *The changing academic profession over 1992–2007: International, comparative, and quantitative perspective*. Hiroshima: RIHE.

Shin, J., & Cummings, W. (2010). Multilevel analysis of academic publishing across disciplines: Research preference, collaboration, and time on research. *Scientometrics, 85*, 581–594.

Smeby, J., & Try, S. (2005). Departmental contexts and faculty research activity in Norway. *Research in Higher Education, 46*(6), 593–619.

Stephan, P., & Levin, S. (1992). *Striking the mother lode in science: The importance of age, place, and time*. Oxford: Oxford University Press.

Stoop, I. (2012). Unit non-response due to refusal. In L. Gideon (Ed.), *Handbook of survey methodology for the social sciences* (pp. 121–147). Dordrecht: Springer.

Teichler, U., Arimoto, A., & Cummings, W. (2013). *The changing academic profession. Major findings of a comparative survey*. Dordrecht: Springer.

Teichler, U., & Höhle, E. (Eds.). (2013). *The work situation of the academic profession in Europe: Findings of a survey in twelve countries*. Dordrecht: Springer.

Teodorescu, D. (2000). Correlates of faculty publication productivity: A cross-national analysis. *Higher Education, 39*, 201–222.

# Appendices (an on-line version only: "Electronic Supplementary Material")

**Table 1**: Characteristics of the sample, by country

|             | Grand N | Universities % | Other HEIs % | Full-time % | Part-time % |
|-------------|---------|----------------|--------------|-------------|-------------|
| Austria     | 1,492   | 100.0          | 0.0          | 65.8        | 34.2        |
| Finland     | 1,374   | 76.5           | 23.5         | 82.4        | 17.6        |
| Germany     | 1,215   | 86.1           | 13.9         | 70.7        | 29.3        |
| Ireland     | 1,126   | 73.3           | 26.7         | 91.2        | 8.8         |
| Italy       | 1,711   | 100.0          | 0.0          | 96.9        | 3.1         |
| Netherlands | 1,209   | 34.4           | 65.6         | 56.0        | 44.0        |
| Norway      | 986     | 93.3           | 6.7          | 89.7        | 10.3        |
| Poland      | 3,704   | 48.3           | 51.7         | 98.0        | 2.0         |
| Portugal    | 1,513   | 40.0           | 60.0         | 90.3        | 9.7         |
| Switzerland | 1,414   | 45.6           | 54.4         | 58.5        | 41.5        |
| UK          | 1,467   | 40.8           | 59.2         | 86.5        | 13.5        |

* In Austria and Italy there was no distinction between "universities" and "other higher education institutions".

**Table 8**: Proportion of faculty by academic field cluster and by country (in percent).

|             | Life sciences, med. sciences | Physical sciences, mathematics | Engineering | Humanities and social sciences | Professions | Other fields | Total |
|-------------|------|------|------|------|------|------|-------|
| Austria     | 20.2 | 9.8  | 11.9 | 41.3 | 8.7  | 8.2  | 1,492 |
| Finland     | 15.7 | 9.7  | 21.5 | 18.6 | 12.1 | 22.4 | 1,374 |
| Germany     | 29.3 | 15.2 | 14.8 | 15.6 | 11.1 | 13.9 | 1,215 |
| Ireland     | 23.0 | 11.5 | 8.8  | 23.8 | 20.5 | 12.4 | 1,126 |
| Italy       | 28.6 | 23.3 | 11.1 | 17.5 | 13.6 | 5.9  | 1,711 |
| Netherlands | 12.6 | 10.9 | 10.7 | 22.3 | 34.7 | 8.8  | 1,209 |
| Norway      | 29.0 | 14.1 | 7.4  | 27.5 | 8.9  | 13.1 | 986   |
| Poland      | 24.6 | 8.4  | 21.5 | 23.0 | 12.5 | 10.0 | 3,704 |
| Portugal    | 16.9 | 7.9  | 20.4 | 10.5 | 20.6 | 23.7 | 1,513 |
| Switzerland | 30.8 | 10.2 | 12.7 | 16.9 | 23.9 | 5.5  | 1,414 |
| UK          | 21.9 | 11.6 | 6.3  | 18.6 | 11.0 | 30.7 | 1,467 |

Note: The major clusters of academic fields are the following: "life sciences and medical sciences" (termed "life sciences" and "medical sciences, health-related sciences, social services" in the survey questionnaire), "physical sciences and mathematics" ("physical sciences, mathematics, computer sciences"), "engineering" ("engineering, manufacturing and construction, architecture"), "humanities and social sciences" ("humanities and arts" and "social and behavioral sciences"), and "professions" ("teacher training and education science", "business and administration, economics", and "law").

**Table 9.** Average research productivity, all items (Question D4 "How many of the following scholarly contributions have you completed in the past <u>three</u> years?" (only academics involved in research, employed full-time, in the university sector).

| Countries/Items | Scholarly books authored or co-authored | Scholarly books edited or co-edited | Articles published in an academic book or journal | Research report monograph written for a funded project | Paper presented at a scholarly conference | A composite country index of research productivity |
|---|---|---|---|---|---|---|
| Austria | 0.6 | 0.7 | 4.3 | 2.1 | 9.5 | 25.3 |
| Finland | 0.4 | 0.4 | 5.5 | 1.4 | 4.9 | 17.6 |
| Germany | 0.4 | 0.4 | 7.8 | 1.9 | 6.7 | 22.6 |
| Ireland | 0.3 | 0.3 | 7.2 | 1.8 | 8.2 | 21.4 |
| Italy | 1.0 | 0.5 | 9.1 | 1.6 | 7.9 | 30.1 |
| Netherlands | 0.5 | 0.6 | 10.7 | 1.7 | 7.6 | 27.4 |
| Norway | 0.5 | 0.2 | 5.0 | 0.7 | 4.5 | 15.0 |
| Poland | 0.2 | 0.2 | 3.9 | 0.2 | 3.2 | 8.8 |
| Portugal | 0.7 | 0.5 | 5.7 | 1.8 | 8.0 | 25.0 |
| Switzerland | 0.6 | 0.4 | 7.8 | 1.7 | 6.1 | 24.2 |
| United Kingdom | 0.3 | 0.2 | 5.9 | 1.1 | 5.7 | 15.8 |
| Item mean | 0.5 | 0.4 | 6.6 | 1.5 | 6.6 | |

Note: the composite country index of research productivity weights particular outputs and aggregates the scores; from among several options of constructing an index e.g. Ramsden 1994: 212-213, Teichler, Arimoto and Cummings 2013: 146-147, and Arimoto 2011: 296, we have selected the latter: we have attributed 10 points for each book, 5 points for an edited book, 1 point for each book chapter or article, and 3 points for each research report.

**Table 10**. Clusters of personal and institutional characteristics linked to individual research productivity, formulations of the relevant survey questions.

*Demographics*
    Female (Question F1): "What is your gender?"
    Mean age (Question F2): Calculated from "Year of birth"
    Married/partner (Question F3): "What is your familial status?"
    Job satisfaction (Question B6): "How would you rate your overall satisfaction with the current job"?
    Father higher education (Question F8): "What is your parents' highest, and if applicable, partners' highest education level?"
    Father primary education (Question F8): What is your parents' highest, and if applicable, partners' highest education level?"
    Full-time (Question A7): "How is your employment situation in the current academic year at your higher education institution/research institute?"
    Professor (Question A10): "What is your academic rank (If you work in a research institutions with ranks differing from those at higher education institutions, please choose the rank most closely corresponding to yours)?"

*Socialization*
    Intensive faculty guidance (Question A3): "How would you characterize the training you received in your doctoral degree? – You received intensive faculty guidance for your research."
    Own research topic (Question A3): "How would you characterize the training you received in your doctoral degree? – You chose your own research topic."
    A scholarship or fellowship (Question A3): "How would you characterize the training you received in your doctoral degree? – You received a scholarship or fellowship."
    Research projects with faculty (Question A3): "How would you characterize the training you received in your doctoral degree? – You were involved in research projects with faculty or senior researchers."

*Internationalization*
    Collaborating internationally (Question D1): "How would you characterize your research efforts undertaken during this (or the previous) academic year? – Do you collaborate with international colleagues?"
    Collaborating domestically (Question D1): "How would you characterize your research efforts undertaken during this (or the previous) academic year? – Do you collaborate with persons at other institutions in your country?"
    Publishing in a foreign country (Question D5): "Which percentage of your publications in the last <u>three</u> years were published in a foreign country?"
    Spending years in other countries (Question F13): "Since the award of your first degree, how many years have you spent in other countries (outside the country of your first degree and current employment)?"
    Research international in scope or orientation (Question D2): "How would you characterize the emphasis of your primary research this (or the previous) academic year?"

*Academic behaviors*
    Mean research hours (in session) (Question B1): "Considering all your professional work, how many hours do you spend in a typical week on each of the following activities?"
    Mean research hours (not in session) (Question B1): "Considering all your professional work, how many hours do you spend in a typical week on each of the following activities?"

*Academic attitudes and role orientation*
    Research-oriented (answer 4 only) (Question B2): "Regarding your own preferences, do your interests lie *primarily* in teaching or in research?"
    Teaching and research compatible (Question B5_7) (inverted): "Please indicate your views on the following: Teaching and research are hardly compatible with each other."
    Research reinforces teaching (Question C4): "Please indicate your views on the following: Your research activities reinforce your teaching."
    Scholarship is original research (Question B5): "Please indicate your views on the following: Scholarship is best defined as the preparation and presentation of findings on original research."

Basic/theoretical research (Question D2): "How would you characterize the emphasis of your primary research this (or the previous) academic year? – Basic/theoretical"

Applied research (Question D2): "How would you characterize the emphasis of your primary research this (or the previous) academic year? – Applied/practically-oriented"

*Overall research engagement*

National/international committees/boards/bodies (Question A13): "During the current academic year, have you done any of the following? – Served as a member of national/international scientific committees/boards/bodies."

A peer reviewer (Question A13): "During the current academic year, have you done any of the following? – Served a peer reviewer (e.g. for journals, research sponsors, institutional evaluations)."

Editor of journals/book series (Question A13): "During the current academic year, have you done any of the following? – Served as an editor of journals/book series."

Local, national or international politics (Question A13): "During the current academic year, have you done any of the following? – Been substantially involved in local, national or international politics."

Technology transfer (Question D3): "Have you been involved in any of the following research activities during this 9or the previous) academic year? – Involved in the process of technology transfer."

Writing research grants (Question D3): "Have you been involved in any of the following research activities during this 9or the previous) academic year? – Answering calls for proposals or writing research grants."

*Institutional policies*

Collegiality (Question E4): "At my institution there is… – Collegiality in decision-making processes."

Strong performance orientation (Question E4): "At my institution there is… – A strong performance orientation."

Performance-based resource allocation (Question E6): "To what extent does your institution emphasize the following practices? – Performance based allocation of resources to academic units."

Commercially-oriented research (Question D6): "Please indicate your views on the following: – Your institution emphasizes commercially-oriented or applied research."

Restrictions on publicly funded research (Question D6): "Please indicate your views on the following: – Restrictions on the publication of results from my publicly-funded research have increased since my first appointment."

Restrictions of privately funded research (Question D6): "Please indicate your views on the following: – Restrictions on the publication of results from my privately-funded research have increased since my first appointment."

Research considered in personnel decisions (Question E6): "To what extent does your institution emphasize the following practices? – Considering the research quality when making personnel decisions."

*Institutional support*

Availability of research funds (Question B3): "At this institution, how would your evaluate each of the following facilities, resources, or personnel you need to support your work? – Research funding."

Availability of research equipment (Question B3): "At this institution, how would your evaluate each of the following facilities, resources, or personnel you need to support your work? – Research equipment and instruments."

Availability of research laboratories (Question B3): "At this institution, how would your evaluate each of the following facilities, resources, or personnel you need to support your work? – Laboratories."

Supportive attitude of administration (Question E4): "At my institution there is… – A supportive attitude of administrative staff towards teaching/research activities."

**Table 11**. Various personal and institutional characteristics linked to individual research productivity, **research top performers (Top) vs. the rest** of academics (**Rest**) (frequencies in percent or averages): patterns holding for *all* countries (universal patterns are shadowed) and for all but one or two countries.

| Items / Countries | AT | | FI | | DE | | IE | | IT | | NL | | NO | | PL | | PT | | CH | | UK | |
|---|---|---|---|---|---|---|---|---|---|---|---|---|---|---|---|---|---|---|---|---|---|---|
| | Top | Rest | Top | Rest | Top | Rest | Top | Rest | Top | Rest | Top | Rest | Top | Rest | Top | Rest | Top | Rest | Top | Rest | Top | Rest |
| Female | 33.8 | 38.7 | 30.8 | 43.0 | 16.2 | 34.1 | 30.5 | 48.9 | 17.8 | 33.9 | 20.8 | 32.7 | 23.9 | 41.2 | 35.9 | 46.0 | 25.7 | 44.0 | 21.4 | 38.4 | 21.5 | 50.1 |
| Mean age | 46.9 | 40.8 | 48.7 | 43.0 | 50.6 | 43.0 | 47.0 | 44.2 | 52.5 | 52.9 | 47.2 | 46.3 | 54.2 | 46.3 | 48.3 | 47.9 | 47.4 | 42.3 | 46.3 | 37.6 | 50.1 | 46.2 |
| Full-time | 82.8 | 66.6 | 91.4 | 81.9 | 91.6 | 71.5 | 95.0 | 92.1 | 96.3 | 96.9 | 85.7 | 63.5 | 95.2 | 89.7 | 98.6 | 97.8 | 95.1 | 90.0 | 77.1 | 57.7 | 97.1 | 85.4 |
| Professor | 8.5 | 4.8 | 44.8 | 8.3 | 34.0 | 5.2 | 32.7 | 6.7 | 43.8 | 29.2 | 41.9 | 8.1 | 69.5 | 18.3 | 16.5 | 6.3 | 9.4 | 1.8 | 31.6 | 6.3 | 40.1 | 8.2 |
| Intensive faculty guidance | 34.6 | 42.2 | 38.5 | 35.9 | 30.1 | 30.7 | 0.0 | 0.0 | 74.8 | 81.3 | 30.9 | 18.5 | 24.3 | 27.9 | 51.4 | 53.5 | 91.9 | 93.1 | 31.3 | 31.5 | 27.8 | 21.7 |
| Research projects with faculty | 41.6 | 41.9 | 64.6 | 51.7 | 39.9 | 37.1 | 0.0 | 0.0 | 76.5 | 66.6 | 24.2 | 15.2 | 36.0 | 43.4 | 42.2 | 44.0 | 67.2 | 51.9 | 52.3 | 48.1 | 45.9 | 39.7 |
| Collaborating internationally | 68.2 | 61.7 | 88.4 | 67.8 | 82.9 | 57.5 | 82.2 | 59.4 | 89.4 | 75.5 | 88.6 | 57.9 | 80.4 | 54.0 | 75.8 | 61.1 | 77.1 | 63.8 | 83.1 | 65.9 | 85.2 | 66.0 |
| Collaborating domestically | 91.7 | 74.4 | 98.7 | 67.8 | 72.4 | 42.8 | 99.0 | 69.7 | 81.2 | 56.6 | 85.7 | 49.4 | 86.8 | 57.4 | 64.4 | 45.0 | 74.9 | 49.1 | 91.0 | 63.3 | 90.8 | 58.0 |
| Publishing in a foreign country | 92.8 | 81.5 | 92.0 | 69.1 | 80.3 | 62.7 | 93.0 | 73.7 | 81.4 | 65.9 | 0.0 | 0.0 | 92.4 | 73.5 | 78.2 | 51.8 | 96.8 | 65.8 | 92.0 | 66.1 | 73.5 | 53.6 |
| Research international in scope | 80.7 | 62.0 | 80.9 | 57.0 | 63.5 | 49.9 | 77.8 | 66.2 | 88.3 | 72.7 | 83.0 | 59.9 | 77.8 | 65.1 | 53.6 | 36.9 | 82.2 | 48.1 | 75.0 | 55.7 | 84.5 | 60.5 |
| Mean research hrs (in session) | 16.9 | 17.8 | 18.3 | 19.6 | 22.1 | 17.0 | 18.2 | 12.8 | 19.6 | 17.1 | 15.9 | 12.9 | 18.1 | 14.6 | 18.7 | 14.4 | 14.6 | 11.2 | 18.2 | 21.4 | 22.7 | 13.8 |
| Mean res. hrs (not in session) | 25.0 | 24.7 | 27.5 | 25.9 | 26.5 | 22.2 | 27.3 | 21.2 | 27.8 | 27.1 | 23.6 | 20.2 | 19.8 | 20.3 | 25.7 | 20.3 | 25.8 | 20.7 | 23.8 | 26.0 | 32.1 | 23.2 |
| Research-oriented (answer 4) | 31.2 | 31.0 | 28.5 | 35.4 | 23.0 | 28.3 | 28.0 | 7.0 | 19.4 | 11.1 | 29.9 | 19.5 | 34.3 | 31.4 | 15.5 | 10.2 | 18.2 | 5.6 | 25.7 | 30.4 | 43.8 | 27.2 |
| Research-oriented (3 & 4) | 82 | 78.6 | 93.9 | 76.1 | 81.4 | 67.5 | 92.0 | 54.4 | 91.4 | 74.6 | 93.5 | 67.3 | 94.6 | 82.6 | 81.9 | 55.7 | 78.0 | 46.8 | 84.5 | 77.6 | 97.0 | 69.8 |
| Research reinforces teaching | 85.3 | 80.4 | 87.3 | 77.7 | 83.7 | 63.4 | 87.4 | 87.2 | 88.6 | 81.7 | 82.4 | 83.2 | 89.6 | 81.2 | 60.1 | 47.9 | 90.4 | 72.2 | 83.3 | 64.6 | 88.7 | 77.6 |
| Scholarship is original research | 81.2 | 71.7 | 76.3 | 60.6 | 80.0 | 68.8 | 73.3 | 71.1 | 78.1 | 72.5 | 83.6 | 75.3 | 95.4 | 87.4 | 74.5 | 67.8 | 84.6 | 69.1 | 0.0 | 0.0 | 79.9 | 68.1 |
| Basic/theoretical research | 72.1 | 70.2 | 72.4 | 55.1 | 64.1 | 57.0 | 64.6 | 48.6 | 61.4 | 56.6 | 54.2 | 51.7 | 69.7 | 67.5 | 62.0 | 57.8 | 50.1 | 41.5 | 46.5 | 43.7 | 69.8 | 52.0 |
| National/internat. committees | 64.8 | 33.7 | 53.1 | 19.1 | 32.5 | 11.3 | 75.0 | 48.9 | 70.2 | 49.9 | 73.0 | 26.7 | 73.2 | 30.3 | 33.8 | 15.4 | 44.5 | 26.1 | 84.7 | 42.0 | 58.2 | 23.3 |
| A peer reviewer | 82.0 | 57.3 | 86.3 | 35.1 | 50.1 | 21.0 | 99.0 | 69.1 | 77.7 | 53.2 | 94.3 | 45.4 | 89.4 | 43.9 | 68.5 | 39.5 | 69.2 | 30.5 | 88.3 | 41.7 | 92.9 | 64.6 |
| Editor of journals/book series | 62.2 | 30.6 | 38.7 | 10.7 | 43.7 | 12.2 | 34.0 | 20.0 | 19.0 | 8.6 | 42.3 | 16.8 | 33.6 | 8.3 | 9.8 | 6.4 | 27.0 | 9.6 | 51.6 | 12.8 | 47.5 | 17.7 |
| Writing research grants | 81.9 | 59.0 | 82.7 | 59.2 | 89.3 | 54.0 | 49.5 | 45.3 | 86.0 | 69.2 | 81.8 | 55.3 | 97.2 | 74.0 | 69.0 | 56.3 | 55.2 | 19.1 | 83.8 | 48.0 | 83.2 | 62.8 |
| Strong performance orientation | 53.4 | 47.0 | 66.3 | 59.9 | 0.0 | 0.0 | 51.1 | 46.4 | 22.2 | 21.8 | 67.1 | 59.6 | 47.4 | 50.9 | 60.2 | 57.3 | 40.3 | 24.2 | 64.0 | 52.7 | 67.5 | 69.6 |
| Research in personnel decisions | 50.2 | 48.7 | 49.6 | 41.9 | 54.0 | 51.0 | 38.6 | 40.2 | 21.8 | 22.6 | 69.8 | 47.4 | 31.2 | 34.6 | 33.9 | 31.8 | 11.5 | 24.1 | 0.0 | 0.0 | 82.2 | 60.8 |
| Availability of research funds | 15.3 | 13.0 | 26.4 | 22.6 | 22.8 | 25.3 | 31.3 | 18.3 | 7.2 | 7.5 | 27.8 | 20.2 | 28.3 | 22.0 | 10.6 | 9.0 | 17.9 | 15.7 | 52.3 | 46.6 | 14.9 | 16.1 |

| Items / Countries | AT | | FI | | DE | | IE | | IT | | NL | | NO | | PL | | PT | | CH | | UK | |
|---|---|---|---|---|---|---|---|---|---|---|---|---|---|---|---|---|---|---|---|---|---|---|
| | Top | Rest | Top | Rest | Top | Rest | Top | Rest | Top | Rest | Top | Rest | Top | Rest | Top | Rest | Top | Rest | Top | Rest | Top | Rest |
| Availability of res. equipment | 52.1 | 45.1 | 59.6 | 53.5 | 54.6 | 52.9 | 64.7 | 54.6 | 36.9 | 29.7 | 58.9 | 38.6 | 39.5 | 52.3 | 36.4 | 34.1 | 36.4 | 30.8 | 72.5 | 70.8 | 44.4 | 38.4 |
| Availability of res. laboratories | 47.5 | 46.1 | 52.8 | 53.6 | 52.9 | 52.4 | 53.5 | 61.6 | 35.0 | 27.8 | 47.7 | 34.3 | 38.0 | 45.4 | 36.9 | 38.4 | 34.4 | 38.3 | 69.1 | 69.4 | 44.2 | 43.4 |
| Supportive attitude of admin. | 39.6 | 30.0 | 20.5 | 24.9 | 30.2 | 22.6 | 49.5 | 46.9 | 18.1 | 16.9 | 38.7 | 29.9 | 33.5 | 36.9 | 21.3 | 22.4 | 14.1 | 12.9 | 50.8 | 54.7 | 34.2 | 33.6 |

Note: In three cases of universal patterns, Italy, "full-time" and "mean age", the Netherlands, "research reinforces teaching", the difference < 1 percent was disregarded.

**Table 12.** Results of t-tests for the equality of means, research top performers (Top) vs. the rest of academics (Rest), all countries. "How long do you spend on various academic activities, only full-time academics in universities involved in research (mean per year, **60 percent when classes are in session and 40 percent** when classes are not in session), hours per week.

| Country | Academic activity | Mean hours per week | | t | p-value | Group with a sig. larger mean (Top or Rest) | % difference | Hours difference per week |
|---|---|---|---|---|---|---|---|---|
| | | Top | Rest | | | | | |
| Austria | Teaching | 10.0 | 8.8 | 1.962 | 0.049 | Top | 13.9 | 1.2 |
| | Research | 19.4 | 20.4 | -0.914 | 0.361 | - | -5.0 | -1.0 |
| | Service | 4.9 | 4.4 | 0.518 | 0.605 | - | 10.5 | 0.5 |
| | Administration | 9.6 | 5.8 | 4.915 | <0.001 | Top | 66.9 | 3.9 |
| | Other | 3.9 | 3.2 | 1.765 | 0.080 | - | 21.3 | 0.7 |
| | Total hours | 46.8 | 42.1 | 3.553 | <0.001 | Top | 11.3 | 4.8 |
| Finland | Teaching | 12.5 | 10.8 | 1.765 | 0.080 | - | 15.3 | 1.7 |
| | Research | 21.4 | 21.2 | 0.103 | 0.918 | - | 0.7 | 0.2 |
| | Service | 2.9 | 2.2 | 1.304 | 0.193 | - | 31.5 | 0.7 |
| | Administration | 6.5 | 4.1 | 3.376 | 0.001 | Top | 58.3 | 2.4 |
| | Other | 3.1 | 2.4 | 1.953 | 0.051 | - | 33.1 | 0.8 |
| | Total hours | 46.5 | 40.9 | 5.058 | <0.001 | Top | 13.5 | 5.5 |
| Germany | Teaching | 10.5 | 9.1 | 1.928 | 0.056 | - | 15.2 | 1.4 |
| | Research | 23.9 | 19.1 | 3.520 | <0.001 | Top | 25.1 | 4.8 |
| | Service | 6.8 | 6.1 | 0.629 | 0.529 | - | 11.9 | 0.7 |
| | Administration | 5.0 | 3.1 | 3.753 | <0.001 | Top | 62.1 | 1.9 |
| | Other | 4.2 | 2.3 | 3.267 | 0.001 | Top | 80.1 | 1.9 |
| | Total hours | 49.7 | 39.5 | 6.166 | <0.001 | Top | 25.9 | 10.2 |
| Ireland | Teaching | 10.5 | 14.5 | -6.438 | <0.001 | Rest | -27.8 | -4.0 |
| | Research | 21.8 | 16.0 | 5.120 | <0.001 | Top | 36.3 | 5.8 |
| | Service | 3.2 | 2.5 | 1.343 | 0.180 | - | 28.5 | 0.7 |
| | Administration | 9.7 | 8.9 | 0.984 | 0.325 | - | 9.1 | 0.8 |
| | Other | 5.3 | 3.7 | 3.129 | 0.002 | Top | 41.4 | 1.6 |
| | Total hours | 50.1 | 45.3 | 3.542 | <0.001 | Top | 10.6 | 4.8 |
| Italy | Teaching | 12.5 | 14.1 | -2.717 | 0.007 | Rest | -11.8 | -1.7 |
| | Research | 22.8 | 21.0 | 2.007 | 0.045 | Top | 8.4 | 1.8 |
| | Service | 4.9 | 3.5 | 1.914 | 0.057 | - | 39.5 | 1.4 |
| | Administration | 4.7 | 4.1 | 1.493 | 0.136 | - | 12.9 | 0.5 |
| | Other | 2.4 | 2.4 | -0.032 | 0.975 | - | -0.4 | 0.0 |
| | Total hours | 47.3 | 45.2 | 2.111 | 0.035 | Top | 4.9 | 2.2 |
| Netherlands | Teaching | 13.9 | 14.5 | -0.410 | 0.682 | - | -3.8 | -0.6 |
| | Research | 19.1 | 15.8 | 2.094 | 0.037 | Top | 20.5 | 3.2 |
| | Service | 2.7 | 2.2 | 1.006 | 0.315 | - | 27.0 | 0.6 |

| Country | Academic activity | Mean hours per week | | t | p-value | Group with a sig. larger mean (Top or Rest) | % difference | Hours difference per week |
|---|---|---|---|---|---|---|---|---|
| | | Top | Rest | | | | | |
| | Administration | 7.7 | 4.3 | 2.626 | 0.011 | Top | 79.4 | 3.4 |
| | Other | 3.3 | 3.0 | 0.385 | 0.701 | - | 8.6 | 0.3 |
| | Total hours | 46.7 | 39.2 | 4.307 | <0.001 | Top | 19.1 | 7.5 |
| Norway | Teaching | 11.2 | 8.1 | 3.896 | <0.001 | Top | 37.3 | 3.0 |
| | Research | 18.8 | 16.9 | 1.571 | 0.117 | - | 11.0 | 1.9 |
| | Service | 2.6 | 1.1 | 3.911 | <0.001 | Top | 142.2 | 1.6 |
| | Administration | 6.2 | 3.6 | 5.306 | <0.001 | Top | 71.2 | 2.6 |
| | Other | 3.3 | 2.0 | 3.226 | 0.002 | Top | 62.2 | 1.3 |
| | Total hours | 40.9 | 31.5 | 5.970 | <0.001 | Top | 30.0 | 9.4 |
| Poland | Teaching | 12.9 | 15.0 | -4.654 | <0.001 | Rest | -14.0 | -2.1 |
| | Research | 20.8 | 16.1 | 6.460 | <0.001 | Top | 28.7 | 4.6 |
| | Service | 3.7 | 3.6 | 0.397 | 0.691 | - | 3.9 | 0.1 |
| | Administration | 5.6 | 4.4 | 2.988 | 0.003 | Top | 28.0 | 1.2 |
| | Other | 3.4 | 3.2 | 0.880 | 0.379 | - | 8.6 | 0.3 |
| | Total hours | 47.7 | 43.5 | 4.582 | <0.001 | Top | 9.5 | 4.1 |
| Portugal | Teaching | 15.4 | 16.0 | -0.682 | 0.495 | - | -3.4 | -0.6 |
| | Research | 19.1 | 14.5 | 4.349 | <0.001 | Top | 31.6 | 4.6 |
| | Service | 1.5 | 1.5 | 0.041 | 0.967 | - | 1.4 | 0.0 |
| | Administration | 5.2 | 4.2 | 1.763 | 0.078 | - | 25.1 | 1.1 |
| | Other | 3.5 | 2.7 | 1.400 | 0.162 | - | 30.5 | 0.8 |
| | Total hours | 44.9 | 39.0 | 3.921 | <0.001 | Top | 15.2 | 5.9 |
| Switzerland | Teaching | 10.1 | 7.2 | 4.140 | <0.001 | Top | 40.8 | 2.9 |
| | Research | 20.3 | 23.2 | -2.867 | 0.005 | Rest | -12.3 | -2.9 |
| | Service | 6.2 | 3.2 | 3.174 | 0.002 | Top | 94.1 | 3.0 |
| | Administration | 7.0 | 5.0 | 3.880 | <0.001 | Top | 39.4 | 2.0 |
| | Other | 4.7 | 3.3 | 2.602 | 0.009 | Top | 42.4 | 1.4 |
| | Total hours | 47.7 | 41.4 | 5.662 | <0.001 | Top | 15.3 | 6.3 |
| United Kingdom | Teaching | 11.3 | 12.8 | -1.402 | 0.164 | - | -11.9 | -1.5 |
| | Research | 25.4 | 17.1 | 5.180 | <0.001 | Top | 48.7 | 8.3 |
| | Service | 2.0 | 1.4 | 1.322 | 0.190 | - | 41.3 | 0.6 |
| | Administration | 9.4 | 9.2 | 0.156 | 0.876 | - | 2.1 | 0.2 |
| | Other | 3.7 | 3.2 | 0.879 | 0.380 | - | 14.9 | 0.5 |
| | Total hours | 50.7 | 43.4 | 4.963 | <0.001 | Top | 16.8 | 7.3 |

9**Table 13**. Results of the z test for the equality of fractions, all countries, preferences for teaching/research (Question B2: "Regarding your own preferences, do your interests lie primarily in teaching or in research?"), **research top performers** (Top) vs. **the rest** of academics (Rest) (percent).

| Country | Teaching-research role orientation | Percent | | Z | p-value | Group with a sig. larger fraction (Top or Rest) |
|---|---|---|---|---|---|---|
| | | Top | Rest | | | |
| Austria | Primarily in teaching | 0.5 | 2.3 | -1.410 | 0.160 | - |
| | In both, but leaning towards teaching | 17.5 | 19.1 | -0.450 | 0.650 | - |
| | In both, but leaning towards research | 50.8 | 47.6 | 0.710 | 0.480 | - |
| | Primarily in research | 31.2 | 31.0 | 0.050 | 0.960 | - |
| Finland | Primarily in teaching | 0.2 | 5.3 | -2.540 | 0.010 | Rest |
| | In both, but leaning towards teaching | 5.9 | 18.6 | -3.550 | <0.001 | Rest |
| | In both, but leaning towards research | 65.5 | 40.7 | 5.240 | <0.001 | Top |
| | Primarily in research | 28.5 | 35.4 | -1.530 | 0.130 | - |
| Germany | Primarily in teaching | 2.6 | 8.6 | -2.190 | 0.030 | Rest |
| | In both, but leaning towards teaching | 16.1 | 23.9 | -1.830 | 0.070 | - |
| | In both, but leaning towards research | 58.4 | 39.2 | 3.850 | <0.001 | Top |
| | Primarily in research | 23.0 | 28.4 | -1.180 | 0.240 | - |
| Ireland | Primarily in teaching | 2.0 | 10.3 | -2.690 | 0.010 | Rest |
| | In both, but leaning towards teaching | 6.0 | 35.2 | -5.900 | <0.001 | Rest |
| | In both, but leaning towards research | 64.0 | 47.4 | 3.120 | <0.001 | Top |
| | Primarily in research | 28.0 | 7.0 | 6.730 | <0.001 | Top |
| Italy | Primarily in teaching | 0.0 | 2.3 | . | . | - |
| | In both, but leaning towards teaching | 8.6 | 23.1 | -4.590 | <0.001 | Rest |
| | In both, but leaning towards research | 72.0 | 63.5 | 2.320 | 0.020 | Top |
| | Primarily in research | 19.4 | 11.1 | 3.310 | <0.001 | Top |
| Netherlands | Primarily in teaching | 0.0 | 6.9 | . | . | - |
| | In both, but leaning towards teaching | 6.5 | 25.7 | -3.340 | <0.001 | Rest |
| | In both, but leaning towards research | 63.6 | 47.8 | 2.320 | 0.020 | Top |
| | Primarily in research | 29.9 | 19.6 | 1.880 | 0.060 | - |
| Norway | Primarily in teaching | 0.0 | 1.1 | . | . | - |
| | In both, but leaning towards teaching | 5.4 | 16.4 | -2.930 | <0.001 | Rest |
| | In both, but leaning towards research | 60.3 | 51.2 | 1.740 | 0.080 | - |
| | Primarily in research | 34.3 | 31.4 | 0.600 | 0.550 | - |
| Poland | Primarily in teaching | 0.7 | 9.4 | -5.930 | <0.001 | Rest |
| | In both, but leaning towards teaching | 17.4 | 34.9 | -7.050 | <0.001 | Rest |
| | In both, but leaning towards research | 66.4 | 45.5 | 7.940 | <0.001 | Top |
| | Primarily in research | 15.5 | 10.2 | 3.210 | <0.001 | Top |
| Portugal | Primarily in teaching | 0.3 | 8.9 | -2.670 | 0.010 | Rest |
| | In both, but leaning towards teaching | 21.7 | 44.3 | -3.860 | <0.001 | Rest |

| Country | Teaching-research role orientation | Percent | | Z | p-value | Group with a sig. larger fraction (Top or Rest) |
|---|---|---|---|---|---|---|
| | | Top | Rest | | | |
| | In both, but leaning towards research | 59.8 | 41.2 | 3.160 | <0.001 | Top |
| | Primarily in research | 18.2 | 5.6 | 4.120 | <0.001 | Top |
| Switzerland | Primarily in teaching | 3.0 | 3.2 | -0.120 | 0.900 | - |
| | In both, but leaning towards teaching | 12.5 | 19.3 | -1.940 | 0.050 | - |
| | In both, but leaning towards research | 58.8 | 47.2 | 2.580 | 0.010 | Top |
| | Primarily in research | 25.7 | 30.4 | -1.130 | 0.260 | - |
| United Kingdom | Primarily in teaching | 0.0 | 4.6 | . | . | - |
| | In both, but leaning towards teaching | 2.9 | 25.7 | -4.620 | <0.001 | Rest |
| | In both, but leaning towards research | 53.2 | 42.6 | 1.840 | 0.070 | - |
| | Primarily in research | 43.8 | 27.2 | 3.140 | <0.001 | Top |

**Table 13.** Odds ratio estimates for logistic regression to be in top 10 percent in research productivity, controlled for country fixed-effects (the reference category being Poland).

| Variable | p-value | Exp(B) |
|---|---|---|
| Nagelkerke's $R^2$ | 0.185 | |
| Cases predicted correctly | 83.80% | |
| Female | 0.064 | 0.826 |
| Mean age | 0.657 | 1.002 |
| Full-time | 0.042 | 1.454* |
| Professor | <0.001 | 1.804*** |
| Intensive faculty guidance | 0.176 | 0.874 |
| Research projects with faculty | 0.062 | 1.192 |
| Collaborating internationally | <0.001 | 1.715*** |
| Collaborating domestically | 0.008 | 1.4** |
| Publishing in a foreign country | 0.004 | 1.47** |
| Research international in scope or orientation | 0.125 | 1.178 |
| Mean research hours (combined: 60 % in session. 40% not in session) | <0.001 | 1.017*** |
| Research-oriented | 0.052 | 1.231 |
| Basic/theoretical research | 0.100 | 0.858 |
| National/international committees/boards/bodies | <0.001 | 1.872*** |
| A peer reviewer | <0.001 | 2.012*** |
| Editor of journals/book series | <0.001 | 1.557*** |
| Availability of research funds | 0.675 | 1.051 |
| Supportive attitude of administration | 0.853 | 0.981 |
| PL | | |
| DE | <0.001 | 2.955*** |
| AT | <0.001 | 3.857*** |
| FI | 0.218 | 1.333 |
| IT | 0.035 | 1.685* |
| NE | <0.001 | 2.764*** |
| NO | <0.001 | 3.444*** |
| PT | 0.580 | 1.148 |
| CH | <0.001 | 2.806*** |
| UK | 0.023 | 1.706* |
| Intercept | <0.001 | 0.005*** |

Note: One variable from the block of academic attitudes and role orientation ("scholarship is original research") and two institutional variables ("strong performance orientation" and "research considered in personal decisions") were not used: there were missing data for one or more countries.

**Table 14.** The distribution of the sample population, by country. A second logistic regression model (controlled for country fixed-effects).

| | Rest | | Top performers | |
|---|---|---|---|---|
| **Country** | **n** | **%** | **n** | **%** |
| Austria | 1,184 | 91.2 | 114 | 8.8 |
| Finland | 977 | 91.9 | 86 | 8.1 |
| Germany | 882 | 87.5 | 125 | 12.5 |
| Ireland | 777 | 89.8 | 88 | 10.2 |
| Italy | 1,323 | 79.0 | 351 | 21.0 |
| Netherlands | 474 | 88.3 | 63 | 11.7 |
| Norway | 817 | 93.2 | 59 | 6.8 |
| Poland | 3,385 | 92.5 | 274 | 7.5 |
| Portugal | 876 | 92.7 | 69 | 7.3 |
| Switzerland | 1,105 | 91.3 | 105 | 8.7 |
| United Kingdom | 715 | 92.0 | 62 | 8.0 |
| Total | 12,515 | 90.0 | 1,396 | 10.0 |